# Femtosecond Laser Writing of Spin Defects in Hexagonal Boron Nitride

Xingyu Gao,[†] Siddhant Pandey,[†] Mehran Kianinia,[‡,¶] Jonghoon Ahn,[§] Peng Ju,[†] Igor Aharonovich,[‡,¶] Niranjan Shivaram,[*,†,∥] and Tongcang Li[*,†,§,⊥,∥]

[†]*Department of Physics and Astronomy, Purdue University, West Lafayette, Indiana 47907, USA*

[‡]*School of Mathematical and Physical Sciences, Faculty of Science, University of Technology Sydney, Ultimo, New South Wales 2007, Australia*

[¶]*ARC Centre of Excellence for Transformative Meta-Optical Systems (TMOS), University of Technology Sydney, Ultimo, New South Wales 2007, Australia*

[§]*School of Electrical and Computer Engineering, Purdue University, West Lafayette, Indiana 47907, USA*

[∥]*Purdue Quantum Science and Engineering Institute, Purdue University, West Lafayette, Indiana 47907, USA*

[⊥]*Birck Nanotechnology Center, Purdue University, West Lafayette, IN 47907, USA*

*E-mail: niranjan@purdue.edu; tcli@purdue.edu

## Abstract

Optically active spin defects in wide-bandgap materials have many potential applications in quantum information and quantum sensing. Spin defects in two-dimensional layered van der Waals materials are just emerging to be investigated. Here we demonstrate that optically-addressable spin ensembles in hexagonal boron nitride (hBN) can be generated by femtosecond laser irradiation. We observe optically detected magnetic resonance (ODMR) of hBN spin defects created by laser irradiation. We show that the creation of spin defects in hBN is strongly affected by the pulse energy of the femtosecond laser. When the laser pulse number is less than a few thousand, the pulse number only affects the density of defects but not the type of defects. With proper laser parameters, spin defects can be generated with a high probability of success. Our work provides a convenient way to create spin defects in hBN by femtosecond laser writing, which shows promising prospects for quantum technologies.

Electron and nuclear spins in solid-state materials offer promising platforms for studying quantum spintronics, and have many applications in quantum sensing and other quantum information technologies.[1–9] The leading candidates such as nitrogen-vacancy (NV) centers in diamond exhibit room-temperature, spin-dependent photon emission that facilitates initialization and readout of spins.[10,11] Inspired by the breakthrough in graphene and other van der Waals materials, a long-sought goal in the field had been to create similar spin qubits in van der Waals materials that can be initialized and read out optically.[12–18]

In this context, hexagonal boron nitride (hBN) is important as it has a wide bandgap[19] and can host high-quality emitters.[20–22] Single-photon emissions above 800 K have been observed with quantum emitters in hBN.[23] Such emitters are widely studied over the past few years due to their outstanding optical properties. They display a high purity of emission, ultrabright count rates, and large stark-shift tuning at room temperature.[24-29] Recent discoveries of negatively charged boron vacancy ($V_B^-$) spin defects in hBN gained broad interest as they display spin-dependent photon emission at room temperature.[15,16] The $V_B^-$ defect consists of a missing boron atom and an extra electron in the hBN crystal. It has been created by high-dose neutron irradiation from a nuclear plant and ion implantation.[15,16] While a few recent works have created defects with a femtosecond laser,[30,31] no spin defects have been observed with laser irradiated hBN samples so far. Here we report the generation of $V_B^-$ defects in hBN using femtosecond laser writing. The creation process is controllable by tuning the laser parameters. With a proper energy of each femotosecond laser pulse and a wide-range of pulse numbers, we can create $V_B^-$ defects with the same emission spectrum as the ones created by neutron and ion implantation.[15,16] Furthermore, the defects generated by laser irradiation exhibit good contrast in optically detected magnetic resonance (ODMR) measurements. Compared with neutron and ion implantation methods,[15,16] our laser irradiation approach is simpler and more flexible as it can be conducted at an ambient environment with no vacuum requirement.[30,32–34] Our work provides a new way to create controllable spin defects in hBN for spin-based quantum technologies.

Figure 1 shows the schematic diagrams of the laser writing process and our confocal system for characterizing samples. Laser writing is done using an 800 nm Ti: Sapphire femtosecond laser with a 50 fs pulse duration and a 1 kHz pulse repetition rate. The Ti: Sapphire femtosecond laser consists of a Coherent Vitara Laser for seeding and a Coherent Elite Duo HE+ amplifier that can generate pulses with a maximum single-pulse energy of 13 mJ. The laser pulse is linearly polarized. It is focused on the hBN sample with an infinity-corrected 50x objective lens with NA = 0.8, which is used to simultaneously image the surface of the hBN sample. The number of pulses incident on the sample is controlled using an optical shutter. The hBN flakes are tape-exfoliated and transferred onto a glass coverslip for laser irradiation. The hBN sample is first placed in the laser focus within the Rayleigh range by monitoring the laser ablation of the hBN surface. The hBN surface is oriented normal to the direction of the laser beam to maintain its position along the laser propagation direction during scanning. For creating defects, the sample is moved 4 μm away from the focus, where the beam waist at the hBN surface is about 3 μm. The high-energy laser pulse breaks chemical bonds in hBN due to multiphoton absorption and creates defects.

After laser irradiation, the samples are moved to our home-built confocal system (Figure 1d) for further characterization, which includes photoluminescence (PL) and ODMR measurements. To investigate the effects of laser parameters on defect generation, we pattern a group of spots. Each spot is irradiated by the femtosecond laser with a different combination of single-pulse energy and total pulse number. The high-energy pulses cause ablation at the center of the laser beam and leave a hole at each laser-written spot (Figure 1c). Because the ablation process is highly nonlinear, the effect of the energy of each 50 fs pulse is very different from the effect of the total energy of many pulses separated by 1 ms.

We carry out a series of measurements to characterize the irradiated spots. We find that within a certain range of pulse energies and pulse numbers of the laser, the spin defects can be generated with a very high probability of success. As shown in Figure 2, at the region irradiated by 20 repetitions of 1 µJ pulses, we observe bright and widely distributed $V_B^-$ defects at the edge of the ablated hole. The confocal image (Figure 2a) of this spot exhibits strong PL emission around the hole. All bright regions in the confocal image have similar emission spectra centered around 820 nm. Fig. 2c shows typical spectra at three different positions (E1, E2, E3) marked in Fig. 2a. The spectra agree with those of spin defects generated by neutron and ion implantation,[15,16] showing a high probability of creating $V_B^-$ defects. In addition, as shown in Fig. 2d, the emitters display good photostability under a 5 mW 532 nm laser excitation over a period of 1000 s. These hBN defects are stable for several months in storage at room temperature.

To further confirm that we have created $V_B^-$ spin defects in hBN by laser irradiation, we perform ODMR measurements of these defects (Fig. 2b). As reported in previous works, the resonant frequencies of the ODMR spectra of $V_B^-$ spin defects are given by $f_\pm = D_{gs} \pm \sqrt{E^2 + \left(\frac{g\mu_B B}{h}\right)^2}$, where $D_{gs}$ = 3.46 GHz and $E$ = 50 MHz are the zero-field splitting parameters of the triplet ground state of $V_B^-$ defects.[15,16] $B$ is the external magnetic field, $g$ is the Landé factor, $\mu_B$ is the Bohr magneton, and $h$ is the Planck constant. To perform the continuous-wave (CW) ODMR measurement, we place a coplanar waveguide close to the laser-irradiated area to apply the microwave. Meanwhile, the sample is excited by a low-power CW 532 nm laser. The emitted photons are counted when the microwave frequency is swept from 3.15 GHz to 3.75 GHz. No external magnetic field is applied. The measured photon counts have two dips at 3.42 GHz and 3.52 GHz, consistent with former ODMR results of $V_B^-$ defects.[15,16] It is noted that the contrast of our measured ODMR can exceed 1 % (Fig. 2b), which is higher than the reported ODMR results of $V_B^-$ defects in hBN so far.[15,16] These show promising spin-optical properties of $V_B^-$ defects created by femtosecond laser writing.

To gain more insight on the formation of $V_B^-$ defects, we pattern a group of laser-writing spots with various pulse energies Es (Figure 3). The energy of each femtosecond pulse is varied from 200 nJ to 1.5 µJ while the pulse number is fixed at 20. The distance between the laser focus and the surface of the hBN flake is 4 µm. A confocal scan followed by the PL emission spectrum measurement is carried out for each spot. Figure 3a shows four typical confocal images of spots created with different pulse energies. In Figure 3b, we compare their PL emission spectra. With different irradiation pulse energies, the intensity and spectrum of PL emission from the resulting spots vary significantly. For low pulse energies (Es ≤ 400 nJ), the PL intensities are very low for most parts of confocal images, indicating the low densities of optically-active defects. Besides, the emission spectra of the bright spots are always shorter than 750 nm, indicating that a different type of defect is created instead of $V_B^-$ spin defects. As the pulse energy increases, the PL intensity increases while the emission spectrum changes as well. At 600 nJ, the emission spectrum varies dramatically at different positions. In this pulse energy region, different emission spectra are observed, including both $V_B^-$ spin defects and other defects. An example emission spectrum for Es = 600 nJ is shown in

Figure 3b. Such emitters with different emission spectra generated by low-energy pulses have no spin signal in the ODMR measurement (Figure 3c). The origin of these different types of emitters is still unknown, but it is out of the scope of this work.

The $V_B^-$ spin defects, which are our main focus, are observed from the sites exposed to pulses with Es ≥ 800 nJ. For defects created by the four different high-energy pulses (800 nJ, 1 µJ, 1.25 µJ and 1.5 µJ), the emission spectra are centered around 820 nm (Fig. 3b) and the ODMR measurements have spin signals (Fig. 3c), which confirm the successful generation of $V_B^-$ defects. Especially, at 1 µJ and 1.25 µJ, we have the highest probability of success to create $V_B^-$ defects.

To get a more general conclusion on the relation between the generation of spin defects, pulse energy, and distance from the laser focus to the sample, we do the laser irradiation on another hBN flake sample with different distances and pulse energies (Fig. 4). In Fig. 4(b), a 2D histogram is plotted to show the dependence of total photon counts from spin defects on the laser pulse energy and the distance between the laser focus and hBN flake. The photon counts are obtained by spectrally and spatially integrating the PL intensity from the confocal scan. For each confocal scan, a 13 $\mu m$ × 13 $\mu m$ area is selected to include the whole region of the hole and the surrounding area. We use an 800 nm longpass filter to block the photons coming from other types of defects. The highest count rate is obtained by using 1 $\mu$J laser pulses with the focus 4 $\mu$m away from the sample. In the high pulse energy regions (≥ 600 nJ), the total counts first increase when we move the laser focus away from the sample's surface and then decreases after reaching maxima (4 $\mu m$ for 1 $\mu$J and 800 nJ, 3 µm for 600 nJ). This is due to the competing effects of an increased irradiation area and a reduced peak laser intensity. In the low pulse energy regions (≥ 400 nJ), the counts keep decreasing when we increase the distance between the laser focus and the sample. In this region, even if the laser is focused on the hBN flake, the photon count is much lower than the high energy region, indicating the low probability of creating spin defects. Few spin defects are observed using an off-focus low-pulse-energy laser (< 200 nJ).

Next, we study the influence of pulse number on the formation of $V_B^-$ defects. A number of laser pulses with 1 μJ single-pulse energy are delivered to write defects at each spot. The pulse number N is varied from 20 to 5000 for different spots. The laser focus is moved 4 μm away from the surface of the hBN flakes. As shown in Fig. 5, the center wavelength of PL spectrum is not sensitive to the pulse number within our experimental range. All these regions exhibit ODMR signals. So the type of defects to be generated is insensitive to the number of pulses. In addition, the PL intensity has no significant change at small pulse numbers (N < 200), while it decreases for very large pulse numbers (N ≥ 1000). We attribute this observation to the annealing effect of a large number of laser pulses. By using high laser energy, a small number of pulses is enough to generate $V_B^-$ defects and cause ablation. The following pulses won't help to create more $V_B^-$ defects because the center part of the flake has been ablated already. However, with a large number of following pulses, the high-energy pulsed laser can lead to an annealing effect, which causes a reduction of photon emission of $V_B^-$ defects as has been reported.[16] In contrast, a pulsed laser has been used for annealing diamond to help create NV color centers.[34] This is because the diffusion of vacancies and nitrogen atoms is required to form a nitrogen-vacancy pair in a diamond. On the other hand, diffusion of boron vacancies in hBN will lead to a combination of them to form more complex defects that do not have spin signals.

Last, we create 20 spots with the same laser irradiation parameters (1 μJ pulse energy and 4 μm between the laser focus and sample) to quantitatively study the probability of generating $V_B^-$ defects under the optimal conditions. As shown in Fig. 6, we do the confocal scan for all 20 spots and calculate the integrated photon counts for each spot. $V_B^-$ defects are stably generated at every spot, while the total photon counts fluctuates with a relative standard deviation around 12.2%.

In conclusion, we have performed controllable laser writing of optically-addressable spin defects in hBN. We show that the single-pulse energy significantly affects the type of defects generated, while the pulse number only affects the density of the created defects. With proper pulse energy (Es ~ 1 μJ) and pulse number (N < 200), $V_B^-$ defects can be generated with high probability of success. Furthermore, the $V_B^-$ defects produced by laser irradiation show a good ODMR contrast at room temperature. This result shows promising spin-optical properties of $V_B^-$ defects for quantum sensing and other applications. Further efforts are needed to non-destructively create $V_B^-$ defects and isolate single emitters. Moreover, coherent control and dynamical decoupling of electron spins are important to explore for future applications of $V_B^-$ defects. Our work provides a new tool for controllable engineering of spin defects in hBN. Such progress motivates more endeavors to explore spin-based quantum technologies, including sensing and quantum information processing in 2D materials.

# Acknowledgement

We thank supports from the Purdue Quantum Science and Engineering Institute (PQ-SEI) seed grant. We thank helpful discussions with Andres E. Llacsahuanga Allcca and Yong P. Chen. IA and MK acknowledge the Australian Research Council (DP180100077, CE200100010) and the Asian Office of Aerospace Research and Development (FA2386-20-1-4014) for the financial support.

# References

(1) Awschalom, D. D.; Bassett, L. C.; Dzurak, A. S.; Hu, E. L.; Petta, J. R. Quantum spintronics: engineering and manipulating atom-like spins in semiconductors. Science 2013, 339, 1174–1179.

(2) Taylor, J.; Cappellaro, P.; Childress, L.; Jiang, L.; Budker, D.; Hemmer, P.; Yacoby, A.; Walsworth, R.; Lukin, M. High-sensitivity diamond magnetometer with nanoscale resolution. Nature Physics 2008, 4, 810–816.

(3) Shi, F.; Zhang, Q.; Wang, P.; Sun, H.; Wang, J.; Rong, X.; Chen, M.; Ju, C.; Rein- hard, F.; Chen, H., et al. Single-protein spin resonance spectroscopy under ambient conditions. Science 2015, 347, 1135–1138.

(4) Cujia, K.; Boss, J. M.; Herb, K.; Zopes, J.; Degen, C. L. Tracking the precession of single nuclear spins by weak measurements. Nature 2019, 571, 230–233.

(5) Casola, F.; van der Sar, T.; Yacoby, A. Probing condensed matter physics with magnetometry based on nitrogen-vacancy centres in diamond. Nature Reviews Materials 2018, 3, 17088.

(6) Wu, Y.; Jelezko, F.; Plenio, M. B.; Weil, T. Diamond quantum devices in biology. Angewandte Chemie International Edition 2016, 55, 6586–6598.

(7) Sipahigil, A.; Evans, R. E.; Sukachev, D. D.; Burek, M. J.; Borregaard, J.; Bhaskar, M. K.; Nguyen, C. T.; Pacheco, J. L.; Atikian, H. A.; Meuwly, C., et al. An integrated diamond nanophotonics platform for quantum-optical networks. Science 2016, 354, 847–850.

(8) Hoang, T. M.; Ahn, J.; Bang, J.; Li, T. Electron spin control of optically levitated nanodiamonds in vacuum. Nature communications 2016, 7, 12250.

(9) Xu, Z.; qi Yin, Z.; Han, Q.; Li, T. Quantum information processing with closely-spaced diamond color centers in strain and magnetic fields [Invited]. Opt. Mater. Express 2019, 9, 4654–4668.

(10) Doherty, M. W.; Manson, N. B.; Delaney, P.; Jelezko, F.; Wrachtrup, J.; Hollen- berg, L. C. The nitrogen-vacancy colour centre in diamond. Physics Reports 2013, 528, 1–45.

(11) Zhou, Y.; Wang, J.; Zhang, X.; Li, K.; Cai, J.; Gao, W. Self-protected thermometry with infrared photons and defect spins in silicon carbide. Physical Review Applied 2017, 8, 044015.

(12) Exarhos, A. L.; Hopper, D. A.; Patel, R. N.; Doherty, M. W.; Bassett, L. C. Magnetic-field-dependent quantum emission in hexagonal boron nitride at room temperature. Nature communications 2019, 10, 222.

(13) Toledo, J.; De Jesus, D.; Kianinia, M.; Leal, A.; Fantini, C.; Cury, L.; S´afar, G.; Aharonovich, I.; Krambrock, K. Electron paramagnetic resonance signature of point defects in neutron-irradiated hexagonal boron nitride. Physical review B 2018, 98, 155203.

(14) Chejanovsky, N.; Mukherjee, A.; Kim, Y.; Denisenko, A.; Finkler, A.; Taniguchi, T.; Watanabe, K.; Dasari, D. B. R.; Smet, J. H.; Wrachtrup, J. Single spin resonance in a van der Waals embedded paramagnetic defect. arXiv preprint arXiv:1906.05903 2019,

(15) Gottscholl, A.; Kianinia, M.; Soltamov, V.; Orlinskii, S.; Mamin, G.; Bradac, C.; Kasper, C.; Krambrock, K.; Sperlich, A.; Toth, M., et al. Initialization and read-out of intrinsic spin defects in a van der Waals crystal at room temperature. Nature Materials 2020, 19, 540–545.

(16) Kianinia, M.; White, S.; Froch, J. E.; Bradac, C.; Aharonovich, I. Generation of spin defects in hexagonal boron nitride. ACS Photonics 2020, 7, 2147–2152.

(17) Gottscholl, A.; Diez, M.; Soltamov, V.; Kasper, C.; Sperlich, A.; Kianinia, M.; Bradac, C.; Aharonovich, I.; Dyakonov, V. Room Temperature Coherent Control of Spin Defects in hexagonal Boron Nitride. arXiv preprint arXiv:2010.12513 2020,

(18) Ivády, V.; Barcza, G.; Thiering, G.; Li, S.; Hamdi, H.; Legeza, Ö.; Chou, J.-P.; Gali, A. Ab initio theory of negatively charged boron vacancy qubit in hBN. arXiv preprint arXiv:1910.07767 2019,

(19) Watanabe, K.; Taniguchi, T.; Kanda, H. Direct-bandgap properties and evidence for ultraviolet lasing of hexagonal boron nitride single crystal. Nature materials 2004, 3, 404–409.

(20) Tran, T. T.; Bray, K.; Ford, M. J.; Toth, M.; Aharonovich, I. Quantum emission from hexagonal boron nitride monolayers. Nature nanotechnology 2016, 11, 37–41.

(21) Tran, T. T.; Elbadawi, C.; Totonjian, D.; Lobo, C. J.; Grosso, G.; Moon, H.; Englund, D. R.; Ford, M. J.; Aharonovich, I.; Toth, M. Robust multicolor single photon emission from point defects in hexagonal boron nitride. ACS nano 2016, 10, 7331–7338.

(22) Ahn, J.; Xu, Z.; Bang, J.; Allcca, A. E. L.; Chen, Y. P.; Li, T. Stable emission and fast optical modulation of quantum emitters in boron nitride nanotubes. Opt. Lett. 2018, 43, 3778–3781.

(23) Kianinia, M.; Regan, B.; Tawfik, S. A.; Tran, T. T.; Ford, M. J.; Aharonovich, I.; Toth, M. Robust solid-state quantum system operating at 800 K. Acs Photonics 2017, 4, 768–773.

(24) Caldwell, J. D.; Aharonovich, I.; Cassabois, G.; Edgar, J. H.; Gil, B.; Basov, D. Photonics with hexagonal boron nitride. Nature Reviews Materials 2019, 4, 552–567.

(25) Novoselov, K.; Mishchenko, o. A.; Carvalho, o. A.; Neto, A. C. 2D materials and van der Waals heterostructures. Science 2016, 353, aac9439.

(26) Liu, W.; Wang, Y.-T.; Li, Z.-P.; Yu, S.; Ke, Z.-J.; Meng, Y.; Tang, J.-S.; Li, C.-F.; Guo, G.-C. Physica E 2020, 124, 114251.

(27) Grosso, G.; Moon, H.; Lienhard, B.; Ali, S.; Efetov, D. K.; Furchi, M. M.; Jarillo-Herrero, P.; Ford, M. J.; Aharonovich, I.; Englund, D. Nat. Commun. 2017, 8, 1-8.

(28) Shotan, Z.; Jayakumar, H.; Considine, C. R.; Mackoit, M.; Fedder, H.; Wrachtrup, J.; Alkauskas, A.; Doherty, M. W.; Menon V. M.; Meriles, C. A. ACS Photon. 2016, 3, 2490-2496.

(29) Xia, Y.; Li, Q.; Kim, J.; Bao, W.; Gong, C.; Yang, S.; Wang, Y.; Zhang, X. Nano Lett. 2019, 19, 7100-7105.

(30) Hou, S.; Birowosuto, M. D.; Umar, S.; Anicet, M. A.; Tay, R. Y.; Coquet, P.; Tay, B. K.; Wang, H.; Teo, E. H. T. Localized emission from laser-irradiated defects in 2D hexagonal boron nitride. 2D Materials 2017, 5, 015010.

(31) Buividas, R.; Aharonovich, I.; Seniutinas, G.; Wang, X.; Rapp, L.; Rode, A. V.; Taniguchi, T.; Juodkazis, S. Photoluminescence from voids created by femtosecond-laser pulses inside cubic-BN. Optics letters 2015, 40, 5711–5713.

(32) Chen, Y.-C.; Salter, P. S.; Knauer, S.; Weng, L.; Frangeskou, A. C.; Stephen, C. J.; Ishmael, S. N.; Dolan, P. R.; Johnson, S.; Green, B. L., et al. Laser writing of coherent colour centres in diamond. Nature Photonics 2017, 11, 77–80.

(33) Castelletto, S.; Almutairi, A.; Kumagai, K.; Katkus, T.; Hayasaki, Y.; Johnson, B.; Juodkazis, S. Photoluminescence in hexagonal silicon carbide by direct femtosecond laser writing. Optics Letters 2018, 43, 6077–6080.

(34) Chen, Y.-C.; Griffiths, B.; Weng, L.; Nicley, S. S.; Ishmael, S. N.; Lekhai, Y.; John- son, S.; Stephen, C. J.; Green, B. L.; Morley, G. W., et al. Laser writing of individual nitrogen-vacancy defects in diamond with near-unity yield. Optica 2019, 6, 662–667.

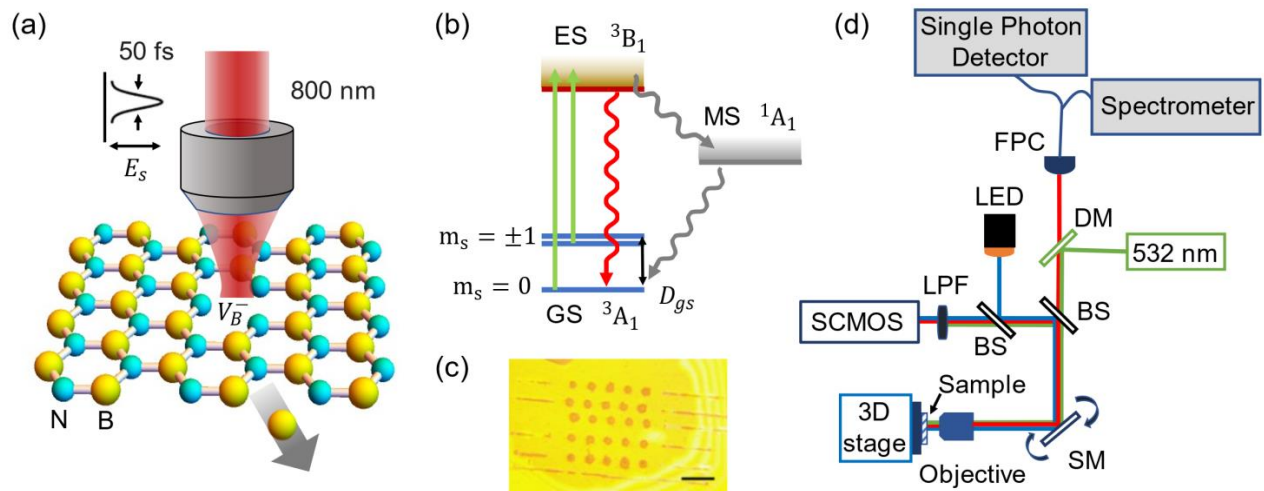

Figure 1: (a) Schematic of the laser writing process. The amplified femtosecond laser is focused by an infinity-corrected 50x objective lens with NA = 0.8. A number of laser pulses are used to generate boron vacancies in hBN flakes. The laser has a center wavelength of 800 nm and a 1 kHz repetition rate. Each laser pulse has a duration of 50 fs. (b) Simplified $V_B^-$ energy-level structure and transitions between the ground state (GS, $^3A_1$), the excited state (GS, $^3B_1$), and the metastable state (MS, $^1A_1$). The ground state has a zero-field splitting $D_{gs}$. (c) An optical image of an hBN flake after femtosecond laser writing. The scale bar is 30 $\mu$m. (d) Simplified schematic of the confocal system (DM: dichroic mirror; SM: scanning mirror; BS: beam splitter; LPF: long pass filter; FPC: fiberport collimator). The laser-irradiated samples are characterized by a separated home-built confocal system. A 532 nm green laser (green lines) is used to excite spin defects in hBN flakes on a substrate. The emitted fluorescence (red lines) is collected with a 100x NA = 0.9 objective lens, and detected by a photon counter and a spectrometer. White light (blue line) from a LED is used to illuminate the sample. The reflected light is detected by a SCMOS camera for imaging the sample.

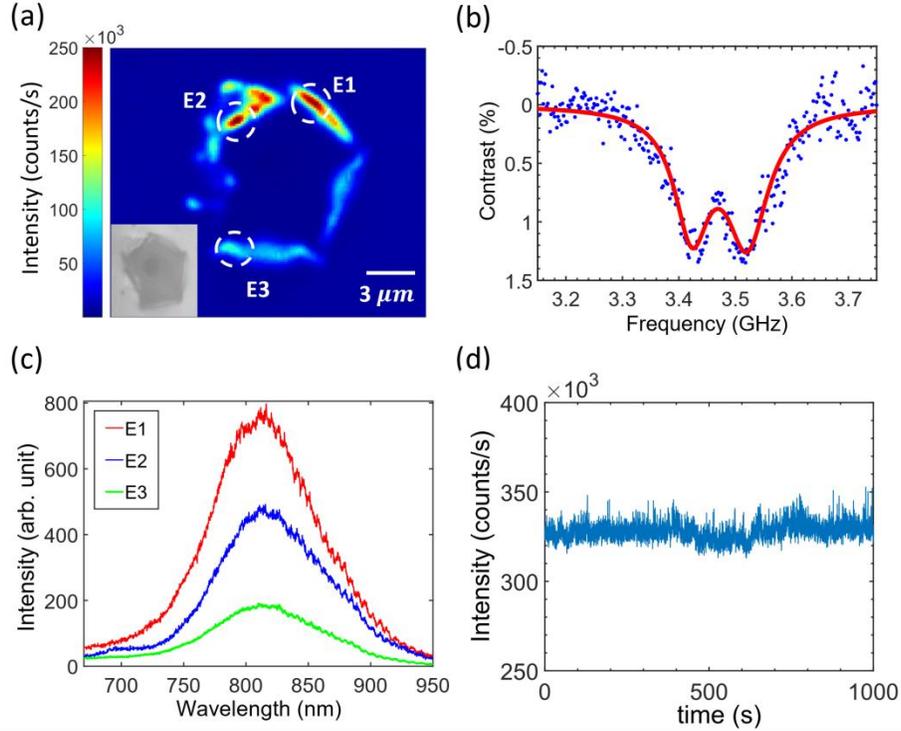

Figure 2: Spin defects in hBN created by femtosecond laser irradiation. (a) A confocal image of a femtosecond-laser irradiated spot. This spot is irradiated by 20 pulses. Each femtosecond pulse has an energy of 1 $\mu$J. The high energy pulses cause ablation of the hBN flake and leave a hole at the center. The spin emitters are observed at the edge of the hole. The emitters are excited by a weak 532 nm laser to obtain the PL image. Inset, a wide-field optical reflection image of the same spot under LED illumination. (b) An ODMR measurement of the spin defects generated by laser irradiation. The defects are excited by a 2.5 mW green laser. The data (blue dots) is fitted by double Lorentzian functions. (c) PL emission spectra of defects marked as E1, E2, E3 in (a). (d) Photostability of $V_B^-$ defects under 5 mW green laser excitation.

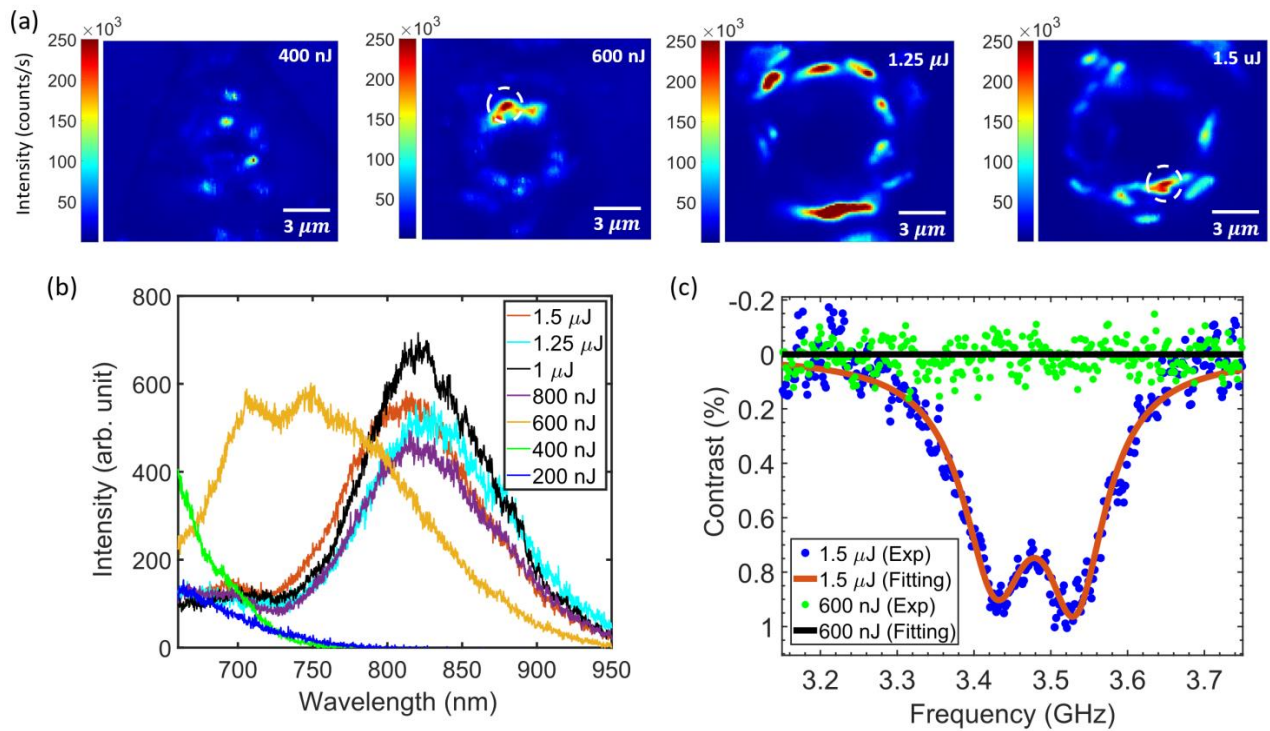

Figure 3: Effects of the pulse energy on defect generation. The pulse energy is varied from 200 nJ to 1.5 μJ in this study. The pulse number is fixed at 20. (a) PL intensity mapping of spots irradiated by lasers with pulse energies $E_s$ = 400 nJ, 600 nJ, 1.25 μJ, and 1.5 μJ. (c) PL emission spectra of emitters under green laser excitation. Spectra of emitters created by laser pulses with pulse energies $E_s$ < 800 nJ show a big difference from the spectrum of $V_B^-$ defects. When $E_s \geq$ 800 nJ, the created defects have similar spectra. (c) ODMR measurements of emitters created by high-energy (1.5 μJ) and low-energy (600 nJ) pulses. The data are obtained from regions marked with white circles in (a). The signal of spin resonance is observed for emitters created by 1.5 μJ pulses while emitters created by 600 nJ pulses do not have spin signals.

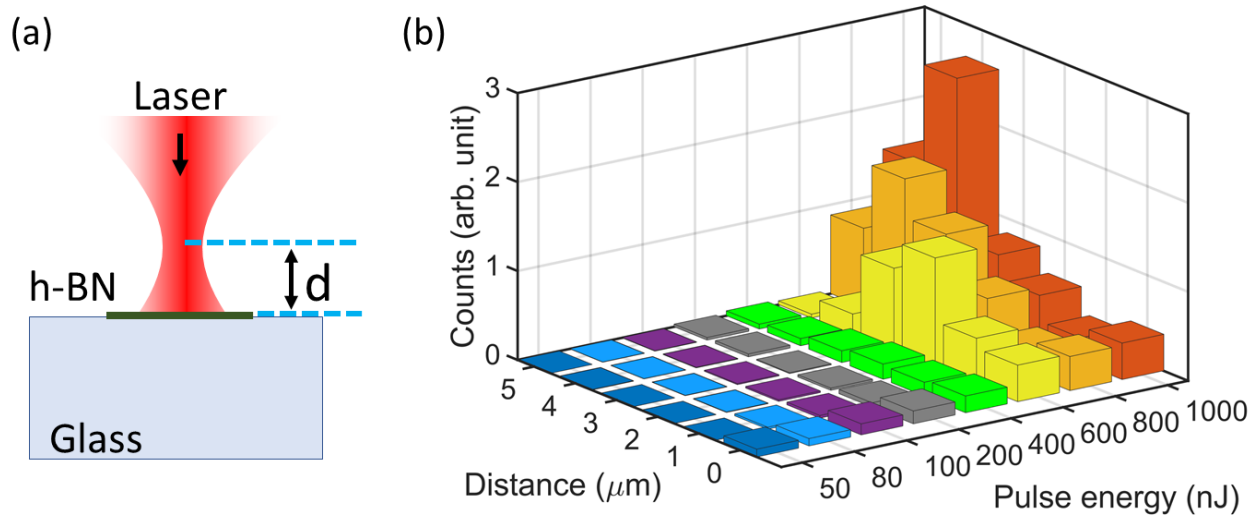

Figure 4: The dependence of the photon counts from the spin defects on the laser pulse energy and the distance between the laser focus and the sample. (a) Schematic of femtosecond laser writing of spin defects in hBN. (b) The spectrally and spatially integrated photon counts. The integrated counts are calculated from the 13 $\mu$m ×13 $\mu$m confocal scan of each laser irradiated spot. An 800 nm longpass filter is used to block the photons from other types of defects.

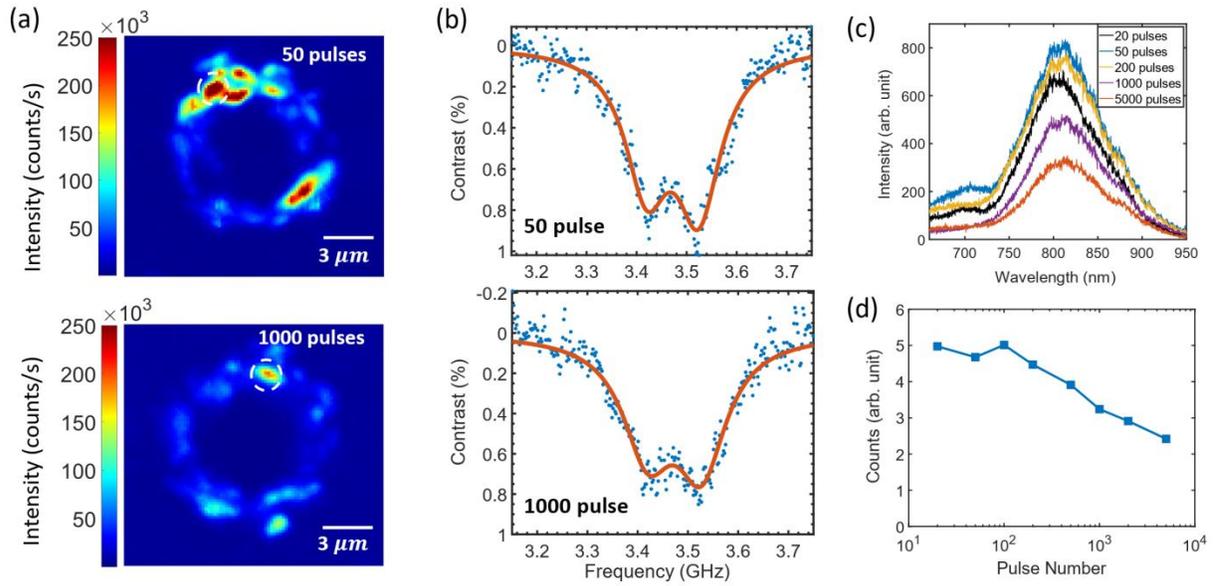

Figure 5: Effects of the pulse number $N$ on defect generation. (a) PL intensity mapping of spots created with different pulse numbers ($N$ = 50 or $N$ = 1000). (b) ODMR measurements of emitters created by 50 pulses or 1000 pulses of fs laser irradiation. Data are obtained from areas marked with white circles in (a). (c) Emission spectra of defects created by laser writing with different pulse numbers. The center wavelength does not change as a function of the pulse number, indicating that the pulse number (within 5000) will not affect the type of defects to be generated. (d) Integrated photon counts as a function of pulse number. The counts are integrated spectrally and spatially from the 13 $\mu$m ×13 $\mu$m confocal scan of each laser irradiated spot.

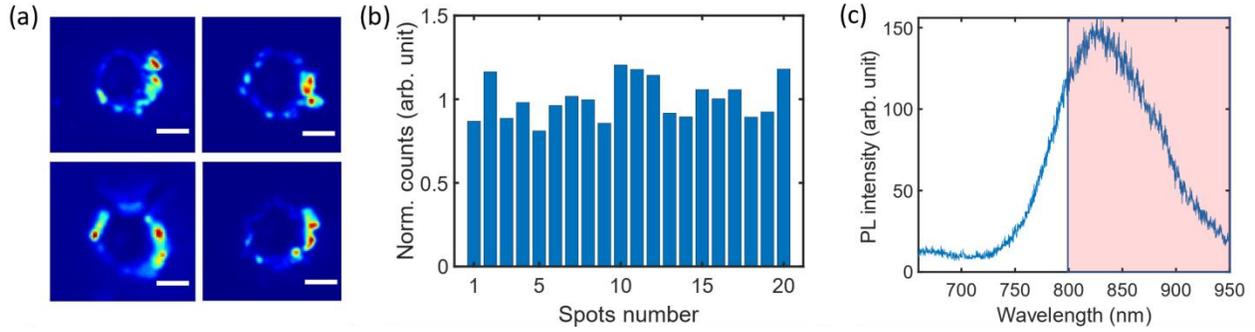

Figure 6: Probability of creating spin defects in hBN by femtosecond laser writing. (a) Confocal images of laser irradiated spots. The size of the confocal scan area is 13 μm × 13 μm. Pulse energy is 1 μJ. The distance between the sample the laser focus is 4 μm. A 800 nm long-pass filter is used to make sure only the photons from $V_B^-$ defects are counted. Scale bar: 3 μm. (b) Spectrally and spatially integrated photon counts normalized by the average count of 20 laser irradiated spots. The integrated spectral range is from 800 nm to 950 nm (colored area as shown in (c)). The relative standard deviation of the integrated counts is 12.2%. (c) Typical spectrum of the bright spots after laser irradiation.